\documentclass[runningheads]{llncs}
\usepackage[T1]{fontenc}
\usepackage{graphicx}
\usepackage{booktabs}
\usepackage{array}
\begin{document}
\title{Synthetic Data Generation for Screen Time and App Usage}

\author{Gustavo Krüger\inst{1,2,3}\orcidID{0000-0002-2771-8595} \and
Nikhil Sachdeva\inst{1,2,3}\orcidID{0000-0001-7544-4552} \and
Michael Sobolev\inst{4}\orcidID{0000-0002-8931-7682}}

\authorrunning{Krüger et al.}

\titlerunning{Screen time synthetic data}
\institute{Universidade Lusófona, Lisbon, Portugal
\and Universitat de Barcelona, Barcelona, Spain
\and Université Paris Cité, Paris, France
\and Cornell Tech, New York, NY, USA \\
\email{gustavo.magro-kruger@etu.u-paris.fr}}

\maketitle

\begin{abstract}
Smartphone usage data can provide valuable insights for understanding interaction with technology and human behavior. However, collecting large-scale, in-the-wild smartphone usage logs is challenging due to high costs, privacy concerns, under representative user samples and biases like non-response that can skew results. These challenges call for exploring alternative approaches to obtain smartphone usage datasets. In this context, large language models (LLMs) such as Open AI's ChatGPT present a novel approach for synthetic smartphone usage data generation, addressing limitations of real-world data collection. We describe a case study on how four prompt strategies influenced the quality of generated smartphone usage data. We contribute with insights on prompt design and measures of data quality, reporting a prompting strategy comparison combining two factors, prompt level of detail (describing a user persona, describing the expected results characteristics) and seed data inclusion (with versus without an initial real usage example). Our findings suggest that using LLMs to generate structured and behaviorally plausible smartphone use datasets is feasible for some use cases, especially when using detailed prompts. Challenges remain in capturing diverse nuances of human behavioral patterns in a single synthetic dataset, and evaluating tradeoffs between data fidelity and diversity, suggesting the need for use-case-specific evaluation metrics and future research with more diverse seed data and different LLM models.

 \and Data Quality
\end{abstract}

\section{Introduction}

Smartphone usage data can provide valuable insights for understanding interaction with technology and human behavior. These data can help the field understand how people use their mobile devices \cite{_ref0,_ref1,_ref2,_ref3,_ref4,_ref5}, infer different behaviors \cite{_ref5.1,_ref5.2}, and also feed into the design of personalized and adaptive interventions \cite{_ref6,_ref7,_ref8,_ref9}. However, collecting large-scale, in-the-wild smartphone usage logs is challenging due to high costs, privacy concerns, under representative user samples, and biases like non-response that can skew results \cite{_ref10,_ref11,_ref12}, justifying the search for alternative approaches to obtain smartphone usage datasets. Synthetic data generation offers an alternative approach that addresses the limitations of real-world data collection \cite{_ref10,_ref11,_ref12}.

Previous work has explored various methods, other than using LLMs, to synthesize smartphone usage data and usage patterns, such as statistical and machine learning models \cite{_ref11,_ref12,_ref13,_ref14}, for example demonstrating that synthetic datasets can mimic real touch interaction patterns and reduce the reliance on extensive field trials \cite{_ref13}. However, these approaches often require training domain-specific generative models, introducing trade-offs in generalization and authenticity \cite{_ref10,_ref11,_ref12,_ref13}.

In this context, Large Language Models (LLMs) such as OpenAI’s ChatGPT present a novel approach for the generation of synthetic smartphone usage data \cite{_ref10,_ref11,_ref12}. LLMs are trained using vast text data that include narratives of technology use, and can be prompted to produce human-like descriptions of user behavior \cite{_ref15,_ref16}. LLMs offer low-cost and low-latency data and the potential for generalization beyond the original scope of a study (e.g. follow-up questions on a survey \cite{_ref15}), which may be especially valuable for hypothesis generation and pilot studies \cite{_ref10,_ref11,_ref12}. For example, Hämäläinen et al. \cite{_ref15} showed that even a relatively outdated LLM like GPT-3 can generate plausible responses in a user experience survey: the model was prompted with the beginning of a real open questions interview about experiencing video games as art, to which it recreated the answers, being able to consistently convince human participants that the responses were human written.

Due to the novelty of this approach, synthetic smartphone usage data generated by LLMs have been scarcely researched, both in terms of how well it reflects actual user behavior \cite{_ref15} and what are possible strategies to generate better data \cite{_ref10,_ref12}, especially with off-the-shelf models. In this case study, we present iterative investigations on the use of the ChatGPT-4o model \cite{_ref17} to generate synthetic datasets of smartphone usage, describing potential benefits and real issues that arise. We describe how four prompt strategies, varying text inputs and the feeding of seed data, influenced the quality of the generated data. Synthetic data was assessed on two levels: (i) structural compliance, which checks that the generated usage logs conform to expected formats and distributions (e.g., event sequences, timing consistency); and (ii) behavioral realism, which evaluates how well the synthetic usage patterns align with real-world usage behaviors (e.g., app session lengths, compatibility with circadian rhythms).

We contribute with insights on prompt design and measures of data quality, reporting a prompting strategy comparison combining two factors, prompt level of detail (describing a user persona, describing the expected results characteristics) and seed data inclusion (with versus without an initial real usage example). We explore how the different prompt designs affect the utility of the generated smartphone interaction data.

\subsection{Prompt Design for Synthetic Data}
Prompt design or engineering became a relevant factor in producing high-quality LLM responses and therefore plays a similar role in generating synthetic data with LLMs. The same LLM can significantly vary its outputs depending upon prompt strategy \cite{_ref18}. Providing examples of the desired data when prompting have been investigated, usually comparing zero-shot, one-shot, or few-shot strategies \cite{_ref16}. In zero-shot prompting, the LLM is given an instruction to generate data without examples \cite{_ref16}. Providing examples is expected to improve realism \cite{_ref18}. Prompt structure and content have also been proposed as relevant variables, as clearly defining the context, format, and constraints of the desired data tend to yield more realistic results \cite{_ref19}. For example, in behavior simulation tasks, explicitly instructing the LLM about roles or formatting (such as delimiting fields like time and location) helps ensure valid and structured outputs \cite{_ref19}.

Another strategy is self-prompting: Reynolds and McDonell (2021) \cite{_ref18} proposed a “MetaPrompt” approach in which an LLM first generates an expanded prompt based on an initial one, which is then used to produce synthetic data. This process is expected to generate better results by leveraging the LLM’s own ability to elaborate on instructions \cite{_ref10,_ref18}, for example, by benefiting from detailed descriptions of the expected data characteristics, known as attribute-controlled prompting \cite{_ref20}, and/or by adding words to the prompt that hold similar semantic meaning in a manner that is relevant to the task, a technique known as verbalizer \cite{_ref21}.

\subsection{Synthetic data evaluation}
Evaluating LLM-generated smartphone usage data requires assessing its generation process, statistical realism and practical utility. The first step is assessing feasibility of data generation, especially for off-the-shelf models, intended for non-expert usage and for which parameters such as temperature (how deterministic or random the LLMs' responses are \cite{_ref16}) or computational power are not controlled by the user \cite{_ref22}. Second, structural compliance with the required format is a prerequisite for other evaluations. Behavioral realism comes last and speaks to the expected usefulness of the synthetic datasets, which should reflect both human compatible patterns and have similar or equivalent distributions, correlations, and temporal patterns as real data. Statistical properties such as app usage frequency, inter-event intervals, and attribute correlations should be preserved for the synthetic data to be meaningful in behavioral modeling \cite{_ref13,_ref15}.

Additionally, the synthetic data literature often discusses the tradeoff between fidelity and novelty, discussing statistics to measure these and compare synthetic and real data \cite{_ref14,_ref19,_ref23}. Another relevant question is to define how data should vary. For example, how should the variety of apps with which users interacts change in a sample? In a LLM generated synthetic dataset, are there new apps that appear? In the current paper, we discuss how variety in an app usage synthetic dataset should be expected in response to the presence or not of seed data in the prompt.

\section{Method}

\subsection{Model choice}
ChatGPT-4o was chosen for the experiment after preliminary testing on free versions of three off-the-shelf LLMs in May of 2025, GTP-4o, DeepSeek and Llama3, employing similar prompts to the ones described in the experiment. DeepSeek and Llama3 presented feasibility issues: DeepSeek usually required more than one subsequent prompt to provide a full version of the dataset, often first outputting only summaries of usage data. It also took approximately four times the time per response of ChatGPT (approximately 4.5 minutes to GPT’s 1.1 minutes). To test Llama3 we used Ollama, a lightweight, extensible framework for building and running language models on a local machine. We encountered two problems 1) since we were running Ollama locally, we were computationally limited and 2) because we only had a text-to-text input-output interaction with the model we could only input the seed data in the prompt text, unlike ChatGPT or DeepSeek that allow for the upload of the seed data as a csv file.

\subsection{Prompting strategy}
We tested four prompts, varying in their level of detail and in the inclusion of seed data. For the more detailed prompts, we employed a self-prompting strategy. Table~\ref{tab:prompt_characteristics} describes the characteristics of each prompt.

\begin{table}[ht]
\small
\caption{Prompts 1 to 4 characteristics.}\label{tab:prompt_characteristics}
\centering
\begin{tabular}{lcc}
\toprule
\textbf{Prompt} & \textbf{Level of detail and origin} & \textbf{Included seed data?} \\
\midrule
P1 & Non-detailed (human designed) & No \\
P2 & Non-detailed (human designed) & Yes \\
P3 & Detailed (self-prompted) & No \\
P4 & Detailed (self-prompted) & Yes \\
\bottomrule
\end{tabular}
\end{table}

Human generated prompts followed current recommended practices \cite{_ref12}, such as asking the model to behave as an expert, asking for a prediction of behavior and clearly describing the desired structure of the data. P1 was the following:

\textit{“You will be asked to predict how people use their smartphones. Your task is to produce a csv file, following a four columns structure. The four columns are: id - identification of the smartphone usage log; timestamp - timestamp for the starting moment of the screen event; app - application under usage; duration - time in seconds of usage. You do not need to start the task now, simply reply with a dot to receive further instructions.”;
“Predict smartphone usage for a day for one person”}

Minimal changes were made for P2 to accommodate for the use of seed data:

\textit{“You will be asked to predict how people use their smartphones. Your task is to produce a csv file, following a four columns structure, according to a model of real usage data for one person, provided in the csv file attached. You do not need to start the task now, simply reply with a dot to receive further instructions.”;
“Predict smartphone usage for the following day for the same person”}

Self-prompting was done using the same model as the synthetic data generation. The following prompt was employed to maximize the level of detail:

\textit{“I plan to generate synthetic data for smartphone usage. the columns of the data look like this - id, created-at, app-id and time-seconds. id is the unique identifier, created-at is the timestamp the session was recorded, app-id is the name of the app used and time-seconds is the time in seconds the app was used for. Now give me a good prompt to generate synthetic data of this format for regular usage on an android device of a young person in their 20s. Add extra context to the prompt, make it as descriptive as possible.”}

The resulting prompt (P3) was the following:
\textit{“You are a data generator tasked with creating realistic synthetic smartphone usage logs for a young adult user in their 20s using an Android smartphone. The data will be used to simulate mobile app usage behavior for research purposes.
Please generate synthetic records in the following format, where each row represents a session in which the user actively used a specific app:
id — a unique identifier for the session (e.g., uuid or simple numeric counter),
created-at — a timestamp in ISO 8601 format (YYYY-MM-DDTHH:MM:SS) when the session started,
app-id — the name of the mobile app used (e.g., Instagram, WhatsApp, Spotify, Chrome),
time-seconds — the number of seconds the app was actively used during this session (integer).
Context and behavioral pattern:
The user is 25 years old, lives in an urban environment, and primarily uses the phone between 8:00 AM and 1:00 AM.
Output Requirements:
Generate data for one full day (24 hours)
Ensure created-at values follow a plausible temporal distribution throughout the day.
Include a mix of app types: social media, messaging, streaming, productivity, utility, and browsing.
Use realistic durations (e.g., 20s for quick WhatsApp check, 1800s for Netflix binge).
Sort the output chronologically by created-at.
Output as a valid CSV file”}

Finally, the same conversation thread was used to generate P4 accommodating for the usage of seed data. The following prompt was added to the thread:

\textit{“Now generate a prompt for generating synthetic data with a seed database”}

Resulting in P4:

\textit{“You are a synthetic data generator tasked with expanding a seed dataset of smartphone app usage logs. The goal is to create additional realistic synthetic records that preserve the behavioral patterns, temporal structure, and app usage characteristics of the original data while not directly copying it.
Schema:
Each row in the dataset represents a session of app usage and contains the following columns:
id: a unique identifier for the session (numeric or UUID; must not repeat any from the seed).
created-at: ISO 8601 timestamp (YYYY-MM-DDTHH:MM:SS) indicating when the app was used.
app-id: name of the app used (e.g., Instagram, WhatsApp, Chrome).
time-seconds: number of seconds the app was used during the session.
Input (Seed Data Overview):
You are given a seed database with many rows, showing real or semi-real app usage from a typical young Android user. The user is in their 20s, lives in an urban environment, and uses apps frequently throughout the day. App categories include social, messaging, entertainment, productivity, and browsing.
Based on this seed:
Learn the behavioral patterns in app usage timing (e.g., WhatsApp use in the morning, Instagram at night),
Preserve statistical patterns of duration (e.g., shorter sessions for messaging, longer for Netflix),
Maintain realistic temporal gaps between sessions (users don’t switch apps every second),
Replicate app frequency trends (e.g., WhatsApp appears more than Medium),
Introduce variation — create new combinations, durations, and timestamps that feel natural but are not copies.
Output Instructions:
Generate new synthetic records as a CSV file with the same 4-column schema.
Use a realistic timestamp progression for a full 24-hour period.
Include realistic patterns
Sort the output chronologically by created-at.
Output must be pure CSV text, no explanation or formatting outside the data.”}

\subsection{Seed data}
Seed data for the prompts P2 and P4 were collected by logging usage of an android smartphone using Android Platform Tools on the device Samsung S20 SM-G781B/DS on the day 17th of April of 2025. Data on the previous and following days (17th and 19th of April) for the same smartphone user were also employed for the behavioral realism analysis metrics B4 and B5, described in detail in the measures section.

\subsection{Procedure}
The experiment was run with the memory settings of ChatGPT turned off and each prompt was tested in a separate chat thread, expecting to have resulting independent synthetic datasets. Two attempts were made with each prompt with reversed orders and on different days. No subsequent prompting was employed and the analyzed datasets were the models' only replies. For attempts with seed data, a csv file was uploaded directly in the chat tread.

\subsection{Measures}
The generated synthetic datasets were subjected to structural and behavioral realism evaluations using the eight metrics in Table~\ref{tab:evaluation_criteria}. The first three metrics (S1, S2 and S3) measure how compliant the synthetic data is to the required format, whereas the subsequent five metrics (B1, B2, B3, B4 and B5) measure how realistic the synthetic data is. Synthetic datasets required a minimal structure compliance in order to be feasible to evaluate their behavioral realism.

\begin{table}[ht]
\small
\caption{Evaluation criteria used to assess structural compliance and behavioral realism of the generated datasets.}
\label{tab:evaluation_criteria}
\centering
\begin{tabular}{>{\raggedright\arraybackslash}p{2.8cm} >{\raggedright\arraybackslash}p{10cm}}
\toprule
\textbf{Criterion} & \textbf{Description} \\
\midrule
\textbf{S1 – Correct variables and formatting} & All required fields (id, timestamp, app\_id, time\_seconds) must be present and use correct formats (e.g., ISO 8601 timestamps). \\
\addlinespace
\textbf{S2 – Raw usage logs as output} & Each row should describe a discrete app usage session. Aggregated summaries or totals are not accepted. \\
\addlinespace
\textbf{S3 – Complete dataset as first output} & The full dataset must be returned in a single model reply without additional prompting. \\
\addlinespace
\textbf{B1 – Realistic daily usage} & Total daily screen time should fall within a realistic range: 1 to 20 hours. \\
\addlinespace
\textbf{B2 – Compatible with circadian rhythm} & At least one continuous non-usage period of 5+ hours should occur during typical sleep hours (between 8 PM and 10 AM). \\
\addlinespace
\textbf{B3 – App variety and usage} & The number and mix of apps used, and their relative usage, should reflect realistic human behavior and resemble the seed data. \\
\addlinespace
\textbf{B4 – Session lengths} & Distribution of session durations (e.g., app use in seconds) should follow plausible behavioral patterns, skewed toward shorter sessions. \\
\addlinespace
\textbf{B5 – Between session intervals} & Gaps between app sessions should show human-like timing, avoiding repetitive or mechanical intervals. \\
\bottomrule
\end{tabular}
\end{table}

The analysis of criteria B4 and B5 were performed comparing the underlying distributions of the synthetic datasets with three days of real usage, one of them being the same one used as seed data. Two methods were employed to group the usage into sessions: 1) consider each log a session, computing log duration distributions and between logs interval’s duration distribution; 2) grouping logs into sessions if they are separated by one minute or more of no smartphone usage.

\subsection{Data availability}
All the datasets utilized in this work, both real (seed data) and synthetic (LLM generated) are available on the project's OSF folder \cite{_ref23.1}.

\section{Results}

\subsection{Structural compliance}

All the attempts to generate synthetic data resulted in a downloadable file with a csv extension, as explicitly required in the prompts. Both attempts for the self-prompts (P3 and P4) resulted in datasets that complied to S1, S2 and S3 criteria. However, the second attempts using P1 and P2 had structural problems that made part of their behavioral realism analysis not feasible - P1 attempt 2 failed to provide raw usages logs, presenting a usage summary instead and P2 attempt 2 provided timestamps that are only dates, not including hours, minutes and seconds.

\subsection{Behavioral realism}
B1 - total usage. Total usage time was within the boundaries of B1 for seven of the synthetic datasets, with the exception of the first attempt of P2, which predicted over 42 hours of usage in a day. Table~\ref{tab:usage_intervals} summarizes the total usage time in hours for each synthetic dataset, as well as the length in hours of their longest interval without any usage.

B2 - Circadian rhythm. Usage predictions were compatible with human circadian rhythm for five of the eight synthetic datasets. Usage was continuos (no intervals) for both P4 attempts and it was not possible to analyze the second attempt of P2 due to no hours and minutes being available on the timestamps column.

\begin{table}[ht]
\small
\caption{Total usage times and length of longest interval without usage for each synthetic dataset.}
\label{tab:usage_intervals}
\centering
\begin{tabular}{lrr}
\toprule
\textbf{Prompt and attempt} & \textbf{Total usage} & \textbf{Longest interval without usage} \\
\midrule
P1.1 & 6,7 & 07:33:17 \\
P1.2 & 2,0 & 08:30:00 \\
P2.1 & 42,3 & 05:41:00 \\
P2.2 & 1,6 & Unable to be verified \\
P3.1 & 6,3 & 07:52:00 \\
P3.2 & 5,8 & 07:40:00 \\
P4.1 & 15,9 & No interval predicted \\
P4.2 & 8,9 & No interval predicted \\
\bottomrule
\end{tabular}
\end{table}

B3 - App usage patterns. App variety was lower in the synthetic datasets if compared to the real data, which describes usage of 33 different applications (Table~\ref{tab:app_overlap}). The two attempts of P4 and the second attempt of P2 yielded the larger pool of apps (18, 16 and 18 respectively), whereas the two attempts of P3 and the first attempt of P2 yielded the smallest pool (10, 9 and 10, respectively).
For prompts with seed data, the five most used apps were exactly the same ones in the seed dataset for all the attempts, the apps being Google Chrome, Google Maps, Lichess, Whatsapp and Instagram, ordered by usage time. Prompts without seed data sometimes yielded different apps as the most used ones, when compared to the real dataset: P1 attempt 1 has all five most used apps different, P1 attempt 2 has four (except for the app “Instagram”), P3 attempt 1 also has four (except app “Google Chrome”) and P3 attempt 2 has two (the same ones are “Google Chrome”, “Google Maps” and “Instagram”). The other most used apps described in P1 and P3 are Camera, Messages, Notes, TikTok, Twitter, Slack, Netflix (two appearances), Spotify (three appearances) and YouTube (three appearances).

\begin{table}[ht]
\small
\caption{App usage patterns for the real and synthetic datasets.}
\label{tab:app_overlap}
\centering
\begin{tabular}{lr>{\raggedleft\arraybackslash}p{5cm}}
\toprule
\textbf{Prompt and attempt} & \textbf{Number of apps} &
\multicolumn{1}{p{5cm}}{\textbf{Percentage of the five most used apps that are the same as the seed data}} \\
\midrule
Real dataset & 33 & -- \\
P1.1 & 15 & 0\% \\
P1.2 & 13 & 20\% \\
P2.1 & 10 & 100\% \\
P2.2 & 18 & 100\% \\
P3.1 & 10 & 20\% \\
P3.2 & 9  & 60\% \\
P4.1 & 18 & 100\% \\
P4.2 & 16 & 100\% \\
\bottomrule
\end{tabular}
\end{table}

\begin{figure}[ht]
\centering
\includegraphics[width=\textwidth]{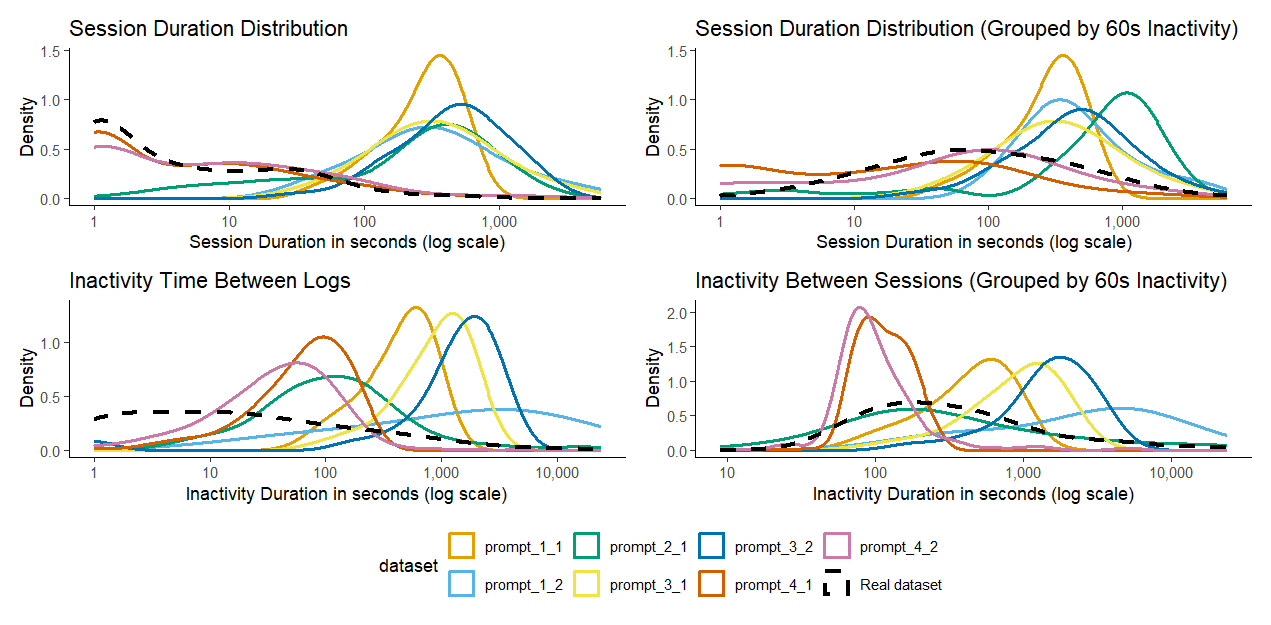}
\caption{Session-level usage and non-usage length distributions for real and synthetic datasets.}
\label{fig:usage_distribution}
\end{figure}

B4 and B5 - realistic session level usage. For session duration, both methods of session grouping yielded similar results: attempts with P4 were the only ones to generate session duration distributions that resemble the real data used as comparisons, characterized by a larger proportion of sessions being relatively shorter (under 100 seconds), whereas P1, P2 and P3 resulted on a larger proportion of longer logs/sessions (Figure~\ref{fig:usage_distribution}). P1 and P3 attempts also were closer to normal distributions, while still skewed to the left (more frequent shorter sessions), whereas P2 resulted in a flatter distribution, centered on longer session durations.

For inactivity intervals, considering time between usage logs, none of the synthetic datasets evaluated resembled the distribution encountered on the real dataset, which is a flatter curve with most intervals ranging from one to 100 seconds. The synthetic datasets, on the other hand, have certain intervals in between usage that happen more frequently, centered around 100 seconds for P2 and P4 attempts and around 1,000 seconds for P3 attempts and for P1 atempt 1. The second attempt of P1 generated a flatter distribution with much larger intervals of no usage (mostly separated by hours - over 3600 seconds). 
When considering time between grouped sessions, the distribution that resulted from P2 resembles the real data, with most intervals ranging from 100 to 300 seconds, whereas the remaining of the synthetic datasets still have more concentrated intervals, around 100 seconds for P4 and around 1,000 seconds for P3 attempts and for P1 attempt 1. The second attempt of P1 generated a flatter distribution with more frequent intervals of no usage that are over one hour long.

\section{Discussion} 

\subsection{Synthetic Data Generation}

Our findings suggest that using LLMs to generate structured and behaviorally plausible smartphone usage datasets is already feasible for some use cases, especially when using detailed prompts, in accordance with previous literature \cite{_ref19,_ref18}. The self-prompting strategies (P3 and P4) consistently generated synthetic datasets according to the required structure, showcasing the potential of LLMs to generate useful data for non-expert users. 

However, for mimicking patterns of real human behavior our results suggest that each proposed use case of synthetic datasets require specific analysis, since none of the tested prompts generated data that satisfied all the proposed behavioral realism criteria. For example, although P4 produced structurally sound data that resembled a real dataset distribution of session durations, it failed to yield any long inactivity intervals, conflicting with known human sleep cycles. This emphasizes that structural fidelity alone or even similarity with one particular measure of realism does not guarantee it for every metric.

We also observed potential effects from procedural variation in prompt order and resource allocation \cite{_ref22}. While not directly measured, discrepancies between repeated prompts suggest that computational factors (e.g., caching, attention allocation) or other unknown sources of variation in generation may influence dataset quality, especially when prompts are tested sequentially, which may be common behavior for non-experts generating synthetic datasets by interacting with chatbots \cite{_ref22}.

\subsection{Challenges of Human-like Simulation}

Despite improvements in structure and plausibility, behavioral realism remains a challenge. Predictions using seed data (P2 and P4) did not consistently align with human-compatible daily cycles. Notably, P4 yielded plausible app usage content but failed the circadian rhythm criteria, highlighting the tension between different criteria for realism. This may reflect a limitation of the LLM in understanding implicit behavioral expectations unless explicitly prompted \cite{_ref18}.

A particularly interesting observation arises from the discrepancy between session-level and log-level analyses. For the first attempt with P2, the resulting dataset's usage intervals did not resemble the real data when comparing log gaps, but session grouping yielded realistic distributions. This suggests a form of equivalence over equality: even if the exact inter-log intervals differ, the broader structure of user sessions can remain compatible with human usage patterns, which may be enough for use cases that will only used grouped data, such as session level analysis. Such findings also point to the need to have diverse behavioral metrics for evaluating synthetic data's usefulness \cite{_ref19,_ref23}.

Observing the app pool for the synthetic datasets, the number of unique apps reveal a limitation of LLMs at generating diversified usage, at least with the tested prompt strategies. This was especially the case for prompts without seed data, which generated daily usage of only 11.75 different apps on average. A similar lack of diversity of LLM generated data has been documented in other fields, such as on a political opinion survey \cite{_ref24} and the money request game \cite{_ref25}.

Regarding the most used apps, we observed a tradeoff between novelty and fidelity. Prompts with seed data resulted in fewer unique apps that reproduced exactly the five most used applications as the seed. Conversely, prompts without seed data introduced a broader mix of apps among the most used ones. These patterns echo findings from prior synthetic data literature \cite{_ref19,_ref23}, where overfitting to seed data can preserve core patterns but reduce variability. In our case, this tradeoff may be desirable or undesirable depending on the intended use: for example, hypothesis generation may benefit from novelty\cite{_ref12}, while behavioral modeling of a specific population or even of a specific user probably should prioritize fidelity \cite{_ref13,_ref15}.

\subsection{Future Directions}

Experimenting with larger and more diverse seed datasets is a relevant next step \cite{_ref11}. Introducing seed data that includes multiple users, temporal spans, or behavioral contexts may improve the model's ability to generalize usage patterns beyond a single individual \cite{_ref12,_ref10}. The inclusion of atypical usage days in the seed data might also provoke the model to simulate less routine behaviors and introduce more variation in app usage.

Future research should also examine the performance of other LLMs, including open-sourced models and/or specialized synthetic data generators, which may offer better behavioral alignment or more controllable generation pipelines. During preliminary comparisons with free versions, ChatGPT-4o showed higher compliance to our prompt generation instructions, but the computational and privacy advantages of local models such as LLaMA3 should be taken into account.

Finally, refining evaluation metrics that are use-case specific is critical \cite{_ref10,_ref15,_ref19,_ref23}. Human-centered modeling requires data that behaves plausibly under simulations and investigating metrics derived from behavioral science (e.g., response times, responses to notifications or behavioral interventions, multitasking patterns, context switching etc) may provide deeper insights into the usefulness of synthetic dataset of smartphone usage.

\medskip

\begin{credits}

\ackname{ This work was funded by the European Union Erasmus Mundus Joint Master Grant no: 101048710. Views and opinions expressed are, however, those of the authors only and do not necessarily reflect those of the European Education and Culture Executive Agency (EACEA).}

\discintname{ The authors declare that they have no competing interests.}
\end{credits}

\end{document}